\documentclass[aps,a4paper,pra,floats,amsfonts]{revtex4}

\usepackage{amsmath}
\usepackage{amsfonts}
\usepackage{amssymb}

\newcommand{\be}{\begin{equation}}
\newcommand{\ee}{\end{equation}}
\newcommand{\bea}{\begin{eqnarray}}
\newcommand{\eea}{\end{eqnarray}}
\newcommand{\bean}{\begin{eqnarray*}}
\newcommand{\eean}{\end{eqnarray*}}
\newcommand{\CASE}[2]{\mbox{$#1\!\otimes\!#2$}}

\begin{document}

  \title{Violation of Equalities in Bipartite Qutrits Systems}
  \author{Hossein Movahhedian}
  \affiliation{Department of Physics, Shahrood University of Technology,
    Seventh Tir Square, Shahrood, Iran}
  \email{hossein_movahhedian@catholic.org}

  \begin{abstract}
    We have recently shown that for the special case of a bipartite system with
    binary inputs and outputs there exist equalities in local theories which
    are violated by quantum theory. The amount of white noise tolerated by
    these equalities are twice that of inequalities. In this paper we will
    first introduce an inequality in bipartite qutrits systems which, if
    non-maximally entangled state is used instead of maximally entangled state,
    is violated more strongly by quantum theory. Hence reproducing the results
    obtained in the literature. We will then prove that our equalities in this
    case are violated by quantum theory too, and they tolerate much more white
    noise than inequalities.
    \begin{center} \today \end{center}
  \end{abstract}

\maketitle

\section{Introduction \label{sec_1}}
  An interesting feature of local theories, discovered by John S. Bell in
  1964~\cite{BEL_J_S_64}, is that for a local theory there exists an inequality
  which is violated by quantum theory. Bell inequalities have attracted more
  attention in recent years because it has been shown by Artur K.
  Ekert~\cite{EKE_A_91} that these inequalities can be used to establish a
  secure quantum key distribution. More interestingly, even if quantum theory
  is not correct, using the violations of Bell inequalities and the
  no-signaling principle it is possible to establish a secure quantum key
  distribution, see~\cite{BHK_05, BKP_06}. Since the work of Bell many
  variations of Bell-type expressions have been introduced. These Bell-type
  expressions are in fact a linear combination of joint probabilities of
  outcomes in an experiment with two or more arms; so that on each arm two or
  more local variable settings are available and there are two or more outcomes
  for each variable setting. An inequality is then a Bell-type expression which
  is bounded by an upper and a lower bound. Among these are CHSH~\cite{CHSH_69}
  and CH~\cite{CH_74}. In this paper we restrict ourselves to experiments with
  two arms and two local variable settings on each arm and designate them by
  \CASE{l_1l_2}{r_1r_2} where $l_1(r_1)$ and $l_2(r_2)$ are the number of
  possible outcomes for the first and the second setting on the left(right) arm
  respectively. Two typical \CASE{22}{22} inequalities are CHSH and CH
  inequalities in which the {\em amount of violation}, i.e. the difference
  between
  the value of a Bell expression according to quantum theory and its (extremum)
  value according to local theories, is $0.41421$ and the {\em tolerance of
  white noise}, i.e. the maximum fraction of white noise admixture for which a
  Bell
  expression stops being violated, is $0.29298$. For the case of \CASE{22}{22}
  inequalities, Arthur Fine~\cite{FIN_A_82} proved that the necessary and
  sufficient condition for the existence of local realistic model is that the
  Bell/CH inequalities hold for the joint probabilities of the experiment.

  In 1982, the experiment of Aspect et~al.~\cite{ADR_82} showed that quantum
  theory is non-local. However, as experiments are not error-free, more
  efficient Bell-type expressions were introduced to test the non-locality of
  quantum theory more precisely, see~\cite{GIS_N_07} for details. The
  \CASE{33}{33} case is the next version of Bell-type inequalities which has
  been studied widely in recent years, see~\cite{KGZMZ_00, CKKZO_01, KKCZO_01,
  DKZ_01, CGLMP_02}. The maximum value of the amount of violation and the
  tolerance of \CASE{33}{33} inequalities are currently predicted to be
  $0.87293$ and $0.30385$ respectively for maximally entangled
  state~\cite{CGLMP_02}.

  Recently based on numerical calculations it has been shown in~\cite{MOV_H_07}
  that in \CASE{22}{22}~case, there exist Bell expressions in local theories
  with exact value, which we called them {\it equalities}, and has been proved
  that these equalities are violated more strongly than inequalities. In fact
  in our equalities the amount of violation and the tolerance of white noise
  admixture are $0.41421$ and $0.58579$ respectively. Here the tolerance is
  twice that of inequalities. However, as shown in~\cite{ADGL_02, AGG_05}, for
  \CASE{33}{33} case an inequality is more resistant to noise if non-maximally
  entangled state is used instead of maximally entangled state. Thus in this
  paper we consider the violation of our equalities using non-maximally
  entangled state. But prior to that we introduce an inequality in
  \CASE{33}{33} case whose amount of violation and its tolerance of white noise
  admixture for non-maximally entangled state are a little more than that of
  maximally entangled state which is in agreement with the results of the above
  papers. Then we prove that there are equalities in \CASE{33}{33} case whose
  tolerance of white noise admixture and violation factor would be much more
  than inequalities if non-maximally entangled state is used.

\section{Bell Expressions and the Experimental Setup\label{sec_2}}
  To verify the non-locality in quantum theory we consider a bipartite
  experiment in which one party, say Alice, performs two measurements
  \mbox{$a \in \{1,2$\}} with the outcomes
  \mbox{$i_1 \in \{0, \cdots, l_1 - 1$\}} and
  \mbox{$i_2 \in \{0, \cdots, l_2 - 1$\}} respectively. Similarly Bob performs
  two measurements \mbox{$b \in \{1,2$\}} with the outcomes
  \mbox{$j_1 \in \{0, \cdots, r_1 - 1$\}} and
  \mbox{$j_2 \in \{0, \cdots, r_2 - 1$\}} respectively. So the
  \mbox{$l_1\times l_2\times r_1\times r_2$} quantities $\gamma_{i_1i_2j_1j_2}$
  designate the double joint probabilities that measurements $a=1$, $a=2$,
  $b=1$ and $b=2$  give outcomes $i_1$, $i_2$, $j_1$, and $j_2$ respectively.
  Obviously
  \be
    \sum_{i_1,i_2,j_1,j_2}\gamma_{i_1i_2j_1j_2}=1 .
    \label{eq_1}
  \ee
  Now the joint probabilities, $P_{ab}^{i_aj_b}$, in terms of these double
  joint probabilities would be
  \be
    P_{ab}^{i_aj_b}=\sum_{i_{a'},j_{b'}} \gamma_{i_ai_{a'}j_bj_{b'}}
    \hspace*{30pt} \mbox{with  $a',b'\in \{1,2\}$, $a'\neq a$ and $b'\neq b$}.
    \label{eq_2}
  \ee
  As has been proved in~\cite{MOV_H_07}, two group of constraints,
  i.e. normalizability of joint probabilities and no-signaling reduces the
  number of independent $P$'s to
   \be
    N_I = (l_1+l_2)(r_1+r_2) - (l_1+l_2+r_1+r_2 - 1) .
       \label{eq_3}
  \ee
  Our numerical calculations show that $N_I$ is 15 and 25 for \CASE{22}{33}
  and \CASE{33}{33} cases respectively, which once again confirms the above
  equation. Note that Eq.~(\ref{eq_3}), as mentioned in~\cite{MOV_H_07},
  is not in agreement with the prediction in~\cite{CO_D_GI_N_04}.
  The reason is that one of the above constraints can be
  written in terms of the others and incidentally, this is not taken into account
  in~\cite{CO_D_GI_N_04}. We have fully discussed this in~\cite{MOV_H_07}.
  Also following the discussion in~\cite{MOV_H_07}, in local
  theories a Bell expression can be written as
  \be
    {\mathbb B} = \sum_{s,t,k,l} \lambda_{stkl}P_{st}^{kl}
    \label{eq_4}
  \ee
  where $\lambda$'s are real (and usually integer) numbers. If we use
  Eq.~(\ref{eq_2}) to write
  $P_{st}^{kl}$ in terms of $\gamma_{i_1i_2j_1j_2}$ we get
  \be
    {\mathbb B} =  \sum_{i_1,i_2,j_1,j_2} (\mu_{i_1i_2j_1j_2} -
                   \nu_{i_1i_2j_1j_2}) \gamma_{i_1i_2j_1j_2} ,
             \hspace{5pt} \mu_{i_1i_2j_1j_2} \neq \nu_{i_1i_2j_1j_2}
    \label{eq_5}
  \ee
  where $\mu$'s and $\nu$'s are non-negative real numbers resulted after
  separating the positive and negative terms. Now the upper bound and the lower
  bound of ${\mathbb B}$ can be written as
  \be
    -d \le {\mathbb B} \le c ,
    \label{eq_6}
  \ee
  where $c(d)$ is the greatest of non-negative real numbers $\mu$'s($\nu$'s).

  The experiment we use is very close to that of Collins~et~al. introduced
  in~\cite{CGLMP_02}, but instead of using a maximally entangled qutrit state
  as they did, we use the general bipartite state
  \be
    |\Psi\rangle = \sum_{j,k=0}^{D-1}C_{jk}|j\rangle_A|k\rangle_B,
      \hspace{10pt} \sum_{j,k=0}^{D-1}|C_{jk}|^2=1 .
    \label{eq_7}
  \ee
  Applying a phase transformation as below
  \bea
    |j\rangle_A & \stackrel{Ph.T.}{\longrightarrow} & e^{2\pi i(j\alpha/D)} |j\rangle_A , \\
    \label{eq_8}
    |k\rangle_B & \stackrel{Ph.T.}{\longrightarrow} & e^{2\pi i(k\beta/D)} |k\rangle_B ,
    \label{eq_9}
  \eea
  and the following discrete Fourier transformation
  \bea
    |j\rangle_A \stackrel{D.F.T.}{\longrightarrow} \frac{1}{\sqrt{D}}
      \sum_{m=0}^{D-1}e^{2\pi i(jm/D)}|m\rangle_A ,  \\
    \label{eq_10}
    |k\rangle_B \stackrel{D.F.T.}{\longrightarrow} \frac{1}{\sqrt{D}}
      \sum_{n=0}^{D-1}e^{-2\pi i(kn/D)}|n\rangle_B ,
    \label{eq_11}
  \eea
  would result the final state as
  \be
    |\Psi\rangle = \frac{1}{D}\sum_{j,k,m,n=0}^{D-1}C_{jk}
      e^{(2\pi i/D)[(\alpha + m)j+(\beta - n)k]}|m\rangle_A |n\rangle_B .
    \label{eq_12}
  \ee
  where $\alpha$ and $\beta$ are real constant numbers to be determined later.
  If Alice and Bob measure on $|m\rangle_A$ and $|n\rangle_B$ respectively,
  then the joint probabilities would be
  \be
    P_{ab}^{mn}=\frac{1}{D^2}\left|\sum_{j,k=0}^{D-1}C_{jk}e^{(2i\pi/D)
      [(\alpha+m)j+(\beta-n)k]}\right|^2 .
    \label{eq_13}
  \ee

\section{Inequalities in \CASE{33}{33} Systems\label{sec_3}}
  For \CASE{33}{33} systems we have listed all joint probabilities, $P$'s,
  in terms of $\gamma$'s in appendix~\ref{app_1}. Let's consider the following
  Bell expression in these systems
  \bea
     \begin{array}{llllllllllll}
     {\mathbb I}
         & = &   & P_{11}^{00}  & - & P_{11}^{01}  & - & P_{11}^{10}
             & - & 2P_{11}^{12} & - & 2P_{11}^{20} \\
         &   & - & P_{11}^{21}  & - & P_{12}^{01}  & - & 2P_{12}^{02}
             & - & P_{12}^{10}  & + & P_{12}^{11}  \\
         &   & - & P_{12}^{21}  & + & P_{12}^{22}  & - & P_{21}^{00}
             & + & P_{21}^{01}  & - & P_{21}^{11}  \\
         &   & + & P_{21}^{12}  & - & P_{21}^{21}  & - & 2P_{21}^{22}
             & - & 2P_{22}^{01} & - & P_{22}^{02} \\
         &   & - & P_{22}^{10}  & - & 2P_{22}^{12} & - & P_{22}^{20}
             & + & P_{22}^{22} .  &   &
     \end{array}
    \label{eq_14}
  \eea

  In appendix~\ref{app_2} we have shown that according to local theories we
  must have $-6 \leq {\mathbb I} \leq 0 $.

  If we define the matrix $T_{mn}$ as
  \be
    T_{mn}=\frac{1}{10}
      \left(
      \begin{array}{ccc}
        \sqrt{38} & 0 & 0 \\
        0 & \sqrt{24} & 0 \\
        0 & 0 & \sqrt{38}
      \end{array}
      \right) ,
    \label{eq_15}
  \ee
  and use the experiment discussed in section~\ref{sec_2}, with $D=3$,
  $C_{m-1,n-1}=T_{mn}$,
  $\alpha=\frac{1}{2}\delta_{a2}$ ($\delta_{xy}$ being Kronecker Delta),
  and $\beta=\frac{1}{4}\delta_{b1}-\frac{1}{4}\delta_{b2}$,
  the value of ${\mathbb I}$ predicted by quantum theory and its tolerance are
  $0.91485$ and $0.31386$ respectively. Hence reproducing the results
  in~\cite{ADGL_02, AGG_05}. These should be compared with the results obtained
  in~\cite{KGZMZ_00, CKKZO_01, KKCZO_01, DKZ_01, CGLMP_02} where, if maximally
  entangled states are used, the amount of violation and the tolerance for
  \CASE{33}{33} (\CASE{55}{55}) case are predicted to be $0.87293$ ($0.91054$)
  and $0.30385$ ($0.31284$) respectively.
  
\section{Equalities in \CASE{33}{33} Systems\label{sec_4}}
  Equalities are built from Bell expressions which satisfy
  Eq.~(\ref{eq_6}) with $d=0$ and $c=1$ which we will call {\it formal
  Bell expressions} from now on. Our numerical calculations show that formal
  Bell expressions in \CASE{32}{22}, \CASE{22}{33}, \CASE{32}{32}, and
  \CASE{33}{33} cases, whether violated by quantum theory or not, are all built
  from \CASE{22}{22} case by dividing one (or more) outcomes to two outcomes.
  As an example, the formal Bell expression for \CASE{32}{22} case is obtained
  from \CASE{22}{22} case simply by dividing one of the outcomes in setting
  labeled $a=1$, on the left arm or Alice's measurements, to two outcomes.
  Similarly the formal Bell expression for \CASE{22}{33} case is obtained from
  \CASE{22}{22} case by dividing one of the outcomes in setting labeled $b=1$
  and one of the outcomes in setting labeled $b=2$, on the right arm or Bob's
  measurements, each to two outcomes. Furthermore, if ${\mathbb E}$ is a formal
  Bell expression then its complement ${\mathbb E}_c$, whose sum with
  ${\mathbb E}$ add up to 1,  can be obtained from ${\mathbb E}$ using one of
  the normalizability conditions,  i.e. $\sum_{i,j}P_{ab}^{ij}=1$. So based on
  these numerical results we conjecture that for higher dimensions the
  derivation of equalities must be similar. With this conjecture two formal
  Bell expressions in \CASE{33}{33} case, which are complements of each other
  and are violated by quantum theory, can be built from \CASE{22}{22} case
  easily. One such formal Bell expression is
  \be
    {\mathbb E} = + P_{11}^{00} + P_{11}^{01} + P_{11}^{10}
      + P_{11}^{11} + P_{12}^{22} + P_{21}^{12} - P_{22}^{12} ,
    \label{eq_16}
  \ee
  and its complement ${\mathbb E}_c$ is
  \be
    {\mathbb E}_c = + P_{11}^{02} + P_{11}^{12} + P_{12}^{20}
      + P_{12}^{21} - P_{21}^{12} + P_{22}^{12} .
    \label{eq_17}
  \ee
  In appendix~\ref{app_3} we have shown directly that
  \be
    |{\mathbb E}| + |{\mathbb E}_c| = 1 .
    \label{eq_18}
  \ee

  It is interesting to note that if ${\mathbb E}$ and ${\mathbb E}_c$ are both
  positive, then, according to normalizability condition, both of them must be
  less than or equal to 1 and none of them would be violated by quantum theory.
  Now again using numerical calculations we have found that for the experiment
  discussed in section~\ref{sec_2}, with
  \be
    T_{mn}=\frac{1}{60}
      \left(
      \begin{array}{ccc}
        \sqrt{1302} & 0 & 0 \\
        i\sqrt{60} & \sqrt{834} & 0 \\
        i\sqrt{60} & i\sqrt{132} & \sqrt{1212}
      \end{array}
      \right) ,
    \label{eq_19}
  \ee
  $D=3$, $C_{m-1,n-1}=T_{mn}$,
  $\alpha=\frac{7}{100}\delta_{a1}+\frac{62}{100}\delta_{a2}$ ($\delta_{xy}$
  being Kronecker Delta), and
  $\beta=\frac{16}{100}\delta_{b1}-\frac{3}{10}\delta_{b2}$, the value of
  ${\mathbb E}_c$ predicted by quantum theory is $-0.14895$. So from
  Eq.s~(\ref{eq_16}) and~(\ref{eq_18}) we conclude that according to local
  theories we must have
  \bea
    |{\mathbb E}| & = & |+ P_{11}^{00} + P_{11}^{01} + P_{11}^{10}
                         + P_{11}^{11} + P_{12}^{22} + P_{21}^{12} - P_{22}^{12}|
                         \nonumber  \\
                  & = & 0.85105 .
    \label{eq_20}
  \eea
  However, for these settings the value of ${\mathbb E}$ predicted by quantum
  theory is $1.14895$.

  Although equalities have not been tested in experiments yet (as far as the
  author is aware), they are testable because, as in the case of inequalites,
  in the testing of an equality the measurement of a Bell expression is
  relevant.

  Now let's consider the tolerance of our equality, i.e. the maximum fraction
  of white noise admixture for which the Bell expression~(\ref{eq_16}) stops
  being violated. The Werner state for three dimensional system is represented
  by
  \be
    \rho=p\frac{\openone}{9} + (1-p)|\Psi\rangle\langle\Psi|
    \label{eq_21} ,
  \ee
  where $\Psi$ is the entangled state as in Eq.~(\ref{eq_7}) and
  $p$ is the tolerance of the Bell expression. For such state the
  tolerance of our equality, i.e. Eq.~(\ref{eq_20}), is $0.50203$ which
  exceeds that of previous results by $0.19818$ and with respect to
  inequality~(\ref{eq_14}) the tolerance is increased by $0.18817$. Note
  that in the presence of white noise the value of the Bell
  expression~(\ref{eq_17}), denoted by ${\mathbb E}_c^{noisy}(p)$, would be
  \be
    {\mathbb E}_c^{noisy}(p) = (m_c-n_c)\frac{p}{9}+(1-p){\mathbb E}_c
    \label{eq_22}
  \ee
  where ${\mathbb E}_c$ is the value of the Bell expression according to
  quantum theory (i.e. for $\rho=|\Psi\rangle\langle\Psi|$) and for the above
  experimental setup it is $-0.14895$, and $m_c (n_c)$ is the number of joint
  probabilities, $P_{ab}^{ij}$, with positive (negative) sign, and from
  Eq.~(\ref{eq_17}), $m_c-n_c = 4$. So for $p=0.50203$ it is readily seen
  from Eq.~(\ref{eq_22}) that ${\mathbb E}_c^{noisy}(p)=+0.14895$ and
  consequently according to local theories ${\mathbb E}^{noisy}(p)$ is still
  $0.85105$.

  The amount of violation of our equality, i.e. the difference between the
  value of the Bell expression~(\ref{eq_16}) according to quantum theory and
  its value according to local theories, is $1.14895-0.85105 = 0.29790$ which
  is much less than that of inequality~(\ref{eq_14}) and the one introduced
  in~\cite{CGLMP_02}. However, one should note that Eq.~(\ref{eq_14})
  contains $24$ $P$'s and its {\it range of violation}, i.e. the difference
  between the upper and the lower bound predicted by local realistic theories,
  denoted by $R$, is $6$. But equality~(\ref{eq_20}) only contains $7$ $P$'s
  and its range of violation is $0.85105$, where we have used the exact value
  of the Bell expression for R which seems to be rational.

  Therefore, we suggest the following generalized definition of violation
  factor, $\eta$, in terms of the amount of violation, $\delta$, and the range
  of violation, $R$:
  \be
    \eta = \frac{\delta + R}{R} .
  \ee
  We have used the range of violation in the above definition because there are
  inequalities with different number of $P$'s but the same range of violation.
  With this definition the violation factor of our equality~(\ref{eq_20}) is
  $1.35004$ while that of inequality~(\ref{eq_14}) and the one
  in~\cite{CGLMP_02} are $1.152475$ and $1.14549$ respectively.

\section{Conclusion\label{sec_5}}
  In this paper we introduced an inequality in \CASE{33}{33} case, i.e. in an
  experiment with two arms, two possible measurements on each arm and three
  possible outcomes for each measurement.
  We showed that if non-maximally entangled states are used, the amount by
  which this inequality is violated and the amount of white noise admixture
  that can be added to a pure state so that it stops violating, i.e. its
  tolerance, is not only more than those in \CASE{33}{33} case with maximally
  entangled state, but also more than those of \CASE{55}{55} case, achieving
  the same results obtained by others.

  Our numerical calculations shows that
  the total number of independent $P$'s in \CASE{22}{33} and \CASE{33}{33}
  cases are $15$ and $25$ respectively which is in agreement with our
  analytical calculations in~\cite{MOV_H_07}. Note that the dimension of
  the space of $P$'s predicted in~\cite{CO_D_GI_N_04} for the above cases are
  $14$ and $24$ respectively.

  Based on our numerical calculations we conjectured how to derive the
  equalities in dimensions higher than two. Then we showed that equalities
  exist in \CASE{33}{33} case and their tolerance of white noise and violation
  factor is much more than inequalities. This increasing of tolerance and
  violation factor in turn make the experiments and any other measurements
  related to non-locality much more easier.

  However, we would like to emphasize that the tolerance of white noise and
  violation factor of equalities in \CASE{22}{22} case are $1.52241$ and
  $0.58579$ respectively (see~\cite{MOV_H_07}), which are more than those of
  \CASE{33}{33} equalities discussed in this paper ($0.50203$, $1.35004$
  respectively). So, in contrast to inequalities, according to our calculations
  the higher the dimension of the system, the lower the efficiency of the
  equalities. Note that for inequalities even for an infinite dimensional
  system the tolerance of white noise admixture and the violation factor
  currently predicted in the literature are $0.32656$ and $1.16164$ which are
  less than those of our \CASE{33}{33} equalities.

\section{Acknowledgments\label{sec_6}}
  We acknowledge the financial support of Shahrood University of Technology
  research Grant~No.~24012.

\appendix

\section{List of Joint Probabilities \label{app_1}}
  The joint probabilities, $P_{ab}^{mn}$, in terms of double joint
  probabilities, $\gamma$'s in \CASE{33}{33} case are:

  \begin{displaymath}
     \begin{array}{lllllllllll}
       P_{11}^{00} = & + & \gamma_{0000} & + & \gamma_{0001} & + & \gamma_{0002} & + & \gamma_{0100} & + & \gamma_{0101} \\
                     & + & \gamma_{0102} & + & \gamma_{0200} & + & \gamma_{0201} & + & \gamma_{0202} &   & \vspace*{3pt} \\
       P_{11}^{01} = & + & \gamma_{0010} & + & \gamma_{0011} & + & \gamma_{0012} & + & \gamma_{0110} & + & \gamma_{0111} \\
                     & + & \gamma_{0112} & + & \gamma_{0210} & + & \gamma_{0211} & + & \gamma_{0212} &   & \vspace*{3pt} \\
       P_{11}^{02} = & + & \gamma_{0020} & + & \gamma_{0021} & + & \gamma_{0022} & + & \gamma_{0120} & + & \gamma_{0121} \\
                     & + & \gamma_{0122} & + & \gamma_{0220} & + & \gamma_{0221} & + & \gamma_{0222} &   & \vspace*{3pt} \\
       P_{11}^{10} = & + & \gamma_{1000} & + & \gamma_{1001} & + & \gamma_{1002} & + & \gamma_{1100} & + & \gamma_{1101} \\
                     & + & \gamma_{1102} & + & \gamma_{1200} & + & \gamma_{1201} & + & \gamma_{1202} &   & \vspace*{3pt} \\
       P_{11}^{11} = & + & \gamma_{1010} & + & \gamma_{1011} & + & \gamma_{1012} & + & \gamma_{1110} & + & \gamma_{1111} \\
                     & + & \gamma_{1112} & + & \gamma_{1210} & + & \gamma_{1211} & + & \gamma_{1212} &   & \vspace*{3pt} \\
       P_{11}^{12} = & + & \gamma_{1020} & + & \gamma_{1021} & + & \gamma_{1022} & + & \gamma_{1120} & + & \gamma_{1121} \\
                     & + & \gamma_{1122} & + & \gamma_{1220} & + & \gamma_{1221} & + & \gamma_{1222} &   & \vspace*{3pt} \\
       P_{11}^{20} = & + & \gamma_{2000} & + & \gamma_{2001} & + & \gamma_{2002} & + & \gamma_{2100} & + & \gamma_{2101} \\
                     & + & \gamma_{2102} & + & \gamma_{2200} & + & \gamma_{2201} & + & \gamma_{2202} &   & \vspace*{3pt} \\
       P_{11}^{21} = & + & \gamma_{2010} & + & \gamma_{2011} & + & \gamma_{2012} & + & \gamma_{2110} & + & \gamma_{2111} \\
                     & + & \gamma_{2112} & + & \gamma_{2210} & + & \gamma_{2211} & + & \gamma_{2212} &   & \vspace*{3pt} \\
       P_{11}^{22} = & + & \gamma_{2020} & + & \gamma_{2021} & + & \gamma_{2022} & + & \gamma_{2120} & + & \gamma_{2121} \\
                     & + & \gamma_{2122} & + & \gamma_{2220} & + & \gamma_{2221} & + & \gamma_{2222} &   & \vspace*{3pt} \\
       P_{12}^{00} = & + & \gamma_{0000} & + & \gamma_{0010} & + & \gamma_{0020} & + & \gamma_{0100} & + & \gamma_{0110} \\
                     & + & \gamma_{0120} & + & \gamma_{0200} & + & \gamma_{0210} & + & \gamma_{0220} &   & \vspace*{3pt} \\
       P_{12}^{01} = & + & \gamma_{0001} & + & \gamma_{0011} & + & \gamma_{0021} & + & \gamma_{0101} & + & \gamma_{0111} \\
                     & + & \gamma_{0121} & + & \gamma_{0201} & + & \gamma_{0211} & + & \gamma_{0221} &   & \vspace*{3pt} \\
       P_{12}^{02} = & + & \gamma_{0002} & + & \gamma_{0012} & + & \gamma_{0022} & + & \gamma_{0102} & + & \gamma_{0112} \\
                     & + & \gamma_{0122} & + & \gamma_{0202} & + & \gamma_{0212} & + & \gamma_{0222} &   & \vspace*{3pt} \\
       P_{12}^{10} = & + & \gamma_{1000} & + & \gamma_{1010} & + & \gamma_{1020} & + & \gamma_{1100} & + & \gamma_{1110} \\
                     & + & \gamma_{1120} & + & \gamma_{1200} & + & \gamma_{1210} & + & \gamma_{1220} &   & \vspace*{3pt} \\
       P_{12}^{11} = & + & \gamma_{1001} & + & \gamma_{1011} & + & \gamma_{1021} & + & \gamma_{1101} & + & \gamma_{1111} \\
                     & + & \gamma_{1121} & + & \gamma_{1201} & + & \gamma_{1211} & + & \gamma_{1221} &   & \vspace*{3pt} \\
       P_{12}^{12} = & + & \gamma_{1002} & + & \gamma_{1012} & + & \gamma_{1022} & + & \gamma_{1102} & + & \gamma_{1112} \\
                     & + & \gamma_{1122} & + & \gamma_{1202} & + & \gamma_{1212} & + & \gamma_{1222} &   & \vspace*{3pt} \\
       P_{12}^{20} = & + & \gamma_{2000} & + & \gamma_{2010} & + & \gamma_{2020} & + & \gamma_{2100} & + & \gamma_{2110} \\
                     & + & \gamma_{2120} & + & \gamma_{2200} & + & \gamma_{2210} & + & \gamma_{2220} &   & \vspace*{3pt} \\
       P_{12}^{21} = & + & \gamma_{2001} & + & \gamma_{2011} & + & \gamma_{2021} & + & \gamma_{2101} & + & \gamma_{2011} \\
                     & + & \gamma_{2121} & + & \gamma_{2201} & + & \gamma_{2211} & + & \gamma_{2221} &   & \vspace*{3pt} \\
       P_{12}^{22} = & + & \gamma_{2002} & + & \gamma_{2012} & + & \gamma_{2022} & + & \gamma_{2102} & + & \gamma_{2112} \\
                     & + & \gamma_{2122} & + & \gamma_{2202} & + & \gamma_{2212} & + & \gamma_{2222} &   & \vspace*{3pt} \\
       P_{21}^{00} = & + & \gamma_{0000} & + & \gamma_{0001} & + & \gamma_{0002} & + & \gamma_{1000} & + & \gamma_{1001} \\
                     & + & \gamma_{1002} & + & \gamma_{2000} & + & \gamma_{2001} & + & \gamma_{2002} &   & \vspace*{3pt} \\
       P_{21}^{01} = & + & \gamma_{0010} & + & \gamma_{0011} & + & \gamma_{0012} & + & \gamma_{1010} & + & \gamma_{1011} \\
                     & + & \gamma_{1012} & + & \gamma_{2010} & + & \gamma_{2011} & + & \gamma_{2012} &   & \vspace*{3pt} \\
       P_{21}^{02} = & + & \gamma_{0020} & + & \gamma_{0021} & + & \gamma_{0022} & + & \gamma_{1020} & + & \gamma_{1021} \\
                     & + & \gamma_{1022} & + & \gamma_{2020} & + & \gamma_{2021} & + & \gamma_{2022} &   & \vspace*{3pt} \\
       P_{21}^{10} = & + & \gamma_{0100} & + & \gamma_{0101} & + & \gamma_{0102} & + & \gamma_{1100} & + & \gamma_{1101} \\
                     & + & \gamma_{1102} & + & \gamma_{2100} & + & \gamma_{2101} & + & \gamma_{2102} &   & \vspace*{3pt} \\
       P_{21}^{11} = & + & \gamma_{0110} & + & \gamma_{0111} & + & \gamma_{0112} & + & \gamma_{1110} & + & \gamma_{1111} \\
                     & + & \gamma_{1112} & + & \gamma_{2110} & + & \gamma_{2111} & + & \gamma_{2112} &   & \vspace*{3pt} \\
       P_{21}^{12} = & + & \gamma_{0120} & + & \gamma_{0121} & + & \gamma_{0122} & + & \gamma_{1120} & + & \gamma_{1121} \\
                     & + & \gamma_{1122} & + & \gamma_{2120} & + & \gamma_{2121} & + & \gamma_{2122} &   & \vspace*{3pt} \\
       P_{21}^{20} = & + & \gamma_{0200} & + & \gamma_{0201} & + & \gamma_{0202} & + & \gamma_{1200} & + & \gamma_{1201} \\
                     & + & \gamma_{1202} & + & \gamma_{2200} & + & \gamma_{2201} & + & \gamma_{2202} &   & \vspace*{3pt} \\
     \end{array}
  \end{displaymath}
  \begin{displaymath}
     \begin{array}{lllllllllll}
       P_{21}^{21} = & + & \gamma_{0210} & + & \gamma_{0211} & + & \gamma_{0212} & + & \gamma_{1210} & + & \gamma_{1211} \\
                     & + & \gamma_{1212} & + & \gamma_{2210} & + & \gamma_{2211} & + & \gamma_{2212} &   & \vspace*{3pt} \\
       P_{21}^{22} = & + & \gamma_{0220} & + & \gamma_{0221} & + & \gamma_{0222} & + & \gamma_{1220} & + & \gamma_{1221} \\
                     & + & \gamma_{1222} & + & \gamma_{2220} & + & \gamma_{2221} & + & \gamma_{2222} &   & \vspace*{3pt} \\
       P_{22}^{00} = & + & \gamma_{0000} & + & \gamma_{0010} & + & \gamma_{0020} & + & \gamma_{1000} & + & \gamma_{1010} \\
                     & + & \gamma_{1020} & + & \gamma_{2000} & + & \gamma_{2010} & + & \gamma_{2020} &   & \vspace*{3pt} \\
       P_{22}^{01} = & + & \gamma_{0001} & + & \gamma_{0011} & + & \gamma_{0021} & + & \gamma_{1001} & + & \gamma_{1011} \\
                     & + & \gamma_{1021} & + & \gamma_{2001} & + & \gamma_{2011} & + & \gamma_{2021} &   & \vspace*{3pt} \\
       P_{22}^{02} = & + & \gamma_{0002} & + & \gamma_{0012} & + & \gamma_{0022} & + & \gamma_{1002} & + & \gamma_{1012} \\
                     & + & \gamma_{1022} & + & \gamma_{2002} & + & \gamma_{2012} & + & \gamma_{2022} &   & \vspace*{3pt} \\
       P_{22}^{10} = & + & \gamma_{0100} & + & \gamma_{0110} & + & \gamma_{0120} & + & \gamma_{1100} & + & \gamma_{1110} \\
                     & + & \gamma_{1120} & + & \gamma_{2100} & + & \gamma_{2110} & + & \gamma_{2120} &   & \vspace*{3pt} \\
       P_{22}^{11} = & + & \gamma_{0101} & + & \gamma_{0111} & + & \gamma_{0121} & + & \gamma_{1101} & + & \gamma_{1111} \\
                     & + & \gamma_{1121} & + & \gamma_{2101} & + & \gamma_{2111} & + & \gamma_{2121} &   & \vspace*{3pt} \\
       P_{22}^{12} = & + & \gamma_{0102} & + & \gamma_{0112} & + & \gamma_{0122} & + & \gamma_{1102} & + & \gamma_{1112} \\
                     & + & \gamma_{1122} & + & \gamma_{2102} & + & \gamma_{2112} & + & \gamma_{2122} &   & \vspace*{3pt} \\
       P_{22}^{20} = & + & \gamma_{0200} & + & \gamma_{0210} & + & \gamma_{0220} & + & \gamma_{1200} & + & \gamma_{1210} \\
                     & + & \gamma_{1220} & + & \gamma_{2200} & + & \gamma_{2210} & + & \gamma_{2220} &   & \vspace*{3pt} \\
       P_{22}^{21} = & + & \gamma_{0201} & + & \gamma_{0211} & + & \gamma_{0221} & + & \gamma_{1201} & + & \gamma_{1211} \\
                     & + & \gamma_{1221} & + & \gamma_{2201} & + & \gamma_{2211} & + & \gamma_{2221} &   & \vspace*{3pt} \\
       P_{22}^{22} = & + & \gamma_{0202} & + & \gamma_{0212} & + & \gamma_{0222} & + & \gamma_{1202} & + & \gamma_{1212} \\
                     & + & \gamma_{1222} & + & \gamma_{2202} & + & \gamma_{2212} & + & \gamma_{2222} . &   & \vspace*{3pt} \\
     \end{array}
  \end{displaymath}

\section{The Range of Violation of Inequality ${\mathbb I}$ \label{app_2}}
  The range of violation of inequality~(\ref{eq_14}), i.e. ${\mathbb I}$, can
  be found easily if we write ${\mathbb I}$ in the form of
  Eq.~(\ref{eq_5}). We have

  \bea
     \begin{array}{llllllllllll}
     {\mathbb I}
         & = & + & P_{11}^{00}  & - & P_{11}^{01}  & - & P_{11}^{10}  & - & 2P_{11}^{12} & - & 2P_{11}^{20} \\
         &   & - & P_{11}^{21}  & - & P_{12}^{01}  & - & 2P_{12}^{02} & - & P_{12}^{10}  & + & P_{12}^{11}  \\
         &   & - & P_{12}^{21}  & + & P_{12}^{22}  & - & P_{21}^{00}  & + & P_{21}^{01}  & - & P_{21}^{11}  \\
         &   & + & P_{21}^{12}  & - & P_{21}^{21}  & - & 2P_{21}^{22} & - & 2P_{22}^{01} & - & P_{22}^{02}  \\
         &   & - & P_{22}^{10}  & - & 2P_{22}^{12} & - & P_{22}^{20}  & + & P_{22}^{22} . &   &
     \end{array} \nonumber
  \eea
  Using $P$'s as defined in appendix~\ref{app_1}, the inequality ${\mathbb I}$
  can be written in terms of $\gamma$'s as below 
  \begin{displaymath}
     \begin{array}{llllllllllll}
     {\mathbb I} =
       & + & \gamma_{0000} & + & \gamma_{0001} & + & \gamma_{0002} & + & \gamma_{0100} & + & \gamma_{0101} \\
       & + & \gamma_{0102} & + & \gamma_{0200} & + & \gamma_{0201} & + & \gamma_{0202} &   & \vspace*{3pt} \\

       & - & \gamma_{0010} & - & \gamma_{0011} & - & \gamma_{0012} & - & \gamma_{0110} & - & \gamma_{0111} \\
       & - & \gamma_{0112} & - & \gamma_{0210} & - & \gamma_{0211} & - & \gamma_{0212} &   & \vspace*{3pt} \\

       & - & \gamma_{1000} & - & \gamma_{1001} & - & \gamma_{1002} & - & \gamma_{1100} & - & \gamma_{1101} \\
       & - & \gamma_{1102} & - & \gamma_{1200} & - & \gamma_{1201} & - & \gamma_{1202} &   & \vspace*{3pt} \\

       & - & 2\gamma_{1020} & - & 2\gamma_{1021} & - & 2\gamma_{1022} & - & 2\gamma_{1120} & - & 2\gamma_{1121} \\
       & - & 2\gamma_{1122} & - & 2\gamma_{1220} & - & 2\gamma_{1221} & - & 2\gamma_{1222} &   & \vspace*{3pt} \\

       & - & 2\gamma_{2000} & - & 2\gamma_{2001} & - & 2\gamma_{2002} & - & 2\gamma_{2100} & - & 2\gamma_{2101} \\
       & - & 2\gamma_{2102} & - & 2\gamma_{2200} & - & 2\gamma_{2201} & - & 2\gamma_{2202} &   & \vspace*{3pt} \\

       & - & \gamma_{2010} & - & \gamma_{2011} & - & \gamma_{2012} & - & \gamma_{2110} & - & \gamma_{2111} \\
       & - & \gamma_{2112} & - & \gamma_{2210} & - & \gamma_{2211} & - & \gamma_{2212} &   & \vspace*{3pt} \\

       & - & \gamma_{0001} & - & \gamma_{0011} & - & \gamma_{0021} & - & \gamma_{0101} & - & \gamma_{0111} \\
       & - & \gamma_{0121} & - & \gamma_{0201} & - & \gamma_{0211} & - & \gamma_{0221} &   & \vspace*{3pt} \\

       & - & 2\gamma_{0002} & - & 2\gamma_{0012} & - & 2\gamma_{0022} & - & 2\gamma_{0102} & - & 2\gamma_{0112} \\
       & - & 2\gamma_{0122} & - & 2\gamma_{0202} & - & 2\gamma_{0212} & - & 2\gamma_{0222} &   & \vspace*{3pt} \\

       & - & \gamma_{1000} & - & \gamma_{1010} & - & \gamma_{1020} & - & \gamma_{1100} & - & \gamma_{1110} \\
       & - & \gamma_{1120} & - & \gamma_{1200} & - & \gamma_{1210} & - & \gamma_{1220} &   & \vspace*{3pt} \\

     \end{array} \nonumber
  \end{displaymath}
  \begin{displaymath}
     \begin{array}{llllllllllll}

       & + & \gamma_{1001} & + & \gamma_{1011} & + & \gamma_{1021} & + & \gamma_{1101} & + & \gamma_{1111} \\
       & + & \gamma_{1121} & + & \gamma_{1201} & + & \gamma_{1211} & + & \gamma_{1221} &   & \vspace*{3pt} \\

       & - & \gamma_{2001} & - & \gamma_{2011} & - & \gamma_{2021} & - & \gamma_{2101} & - & \gamma_{2011} \\
       & - & \gamma_{2121} & - & \gamma_{2201} & - & \gamma_{2211} & - & \gamma_{2221} &   & \vspace*{3pt} \\

       & + & \gamma_{2002} & + & \gamma_{2012} & + & \gamma_{2022} & + & \gamma_{2102} & + & \gamma_{2112} \\
       & + & \gamma_{2122} & + & \gamma_{2202} & + & \gamma_{2212} & + & \gamma_{2222} &   & \vspace*{3pt} \\

       & - & \gamma_{0000} & - & \gamma_{0001} & - & \gamma_{0002} & - & \gamma_{1000} & - & \gamma_{1001} \\
       & - & \gamma_{1002} & - & \gamma_{2000} & - & \gamma_{2001} & - & \gamma_{2002} &   & \vspace*{3pt} \\

       & + & \gamma_{0010} & + & \gamma_{0011} & + & \gamma_{0012} & + & \gamma_{1010} & + & \gamma_{1011} \\
       & + & \gamma_{1012} & + & \gamma_{2010} & + & \gamma_{2011} & + & \gamma_{2012} &   & \vspace*{3pt} \\

       & - & \gamma_{0110} & - & \gamma_{0111} & - & \gamma_{0112} & - & \gamma_{1110} & - & \gamma_{1111} \\
       & - & \gamma_{1112} & - & \gamma_{2110} & - & \gamma_{2111} & - & \gamma_{2112} &   & \vspace*{3pt} \\

       & + & \gamma_{0120} & + & \gamma_{0121} & + & \gamma_{0122} & + & \gamma_{1120} & + & \gamma_{1121} \\
       & + & \gamma_{1122} & + & \gamma_{2120} & + & \gamma_{2121} & + & \gamma_{2122} &   & \vspace*{3pt} \\

       & - & \gamma_{0210} & - & \gamma_{0211} & - & \gamma_{0212} & - & \gamma_{1210} & - & \gamma_{1211} \\
       & - & \gamma_{1212} & - & \gamma_{2210} & - & \gamma_{2211} & - & \gamma_{2212} &   & \vspace*{3pt} \\

       & - & 2\gamma_{0220} & - & 2\gamma_{0221} & - & 2\gamma_{0222} & - & 2\gamma_{1220} & - & 2\gamma_{1221} \\
       & - & 2\gamma_{1222} & - & 2\gamma_{2220} & - & 2\gamma_{2221} & - & 2\gamma_{2222} &   & \vspace*{3pt} \\

       & - & 2\gamma_{0001} & - & 2\gamma_{0011} & - & 2\gamma_{0021} & - & 2\gamma_{1001} & - & 2\gamma_{1011} \\
       & - & 2\gamma_{1021} & - & 2\gamma_{2001} & - & 2\gamma_{2011} & - & 2\gamma_{2021} &   & \vspace*{3pt} \\

       & - & \gamma_{0002} & - & \gamma_{0012} & - & \gamma_{0022} & - & \gamma_{1002} & - & \gamma_{1012} \\
       & - & \gamma_{1022} & - & \gamma_{2002} & - & \gamma_{2012} & - & \gamma_{2022} &   & \vspace*{3pt} \\

       & - & \gamma_{0100} & - & \gamma_{0110} & - & \gamma_{0120} & - & \gamma_{1100} & - & \gamma_{1110} \\
       & - & \gamma_{1120} & - & \gamma_{2100} & - & \gamma_{2110} & - & \gamma_{2120} &   & \vspace*{3pt} \\

       & - & 2\gamma_{0102} & - & 2\gamma_{0112} & - & 2\gamma_{0122} & - & 2\gamma_{1102} & - & 2\gamma_{1112} \\
       & - & 2\gamma_{1122} & - & 2\gamma_{2102} & - & 2\gamma_{2112} & - & 2\gamma_{2122} &   & \vspace*{3pt} \\

       & - & \gamma_{0200} & - & \gamma_{0210} & - & \gamma_{0220} & - & \gamma_{1200} & - & \gamma_{1210} \\
       & - & \gamma_{1220} & - & \gamma_{2200} & - & \gamma_{2210} & - & \gamma_{2220} &   & \vspace*{3pt} \\

       & + & \gamma_{0202} & + & \gamma_{0212} & + & \gamma_{0222} & + & \gamma_{1202} & + & \gamma_{1212} \\
       & + & \gamma_{1222} & + & \gamma_{2202} & + & \gamma_{2212} & + & \gamma_{2222} . &   & \vspace*{3pt} \\

     \end{array} \nonumber
  \end{displaymath}
  Simplifying the above equation yields

  \begin{displaymath}
     \begin{array}{llllllllllll}
     {\mathbb I} =
       & - & 3\gamma_{0001} & - & 3\gamma_{0002} & - & 3\gamma_{0011} & - & 3\gamma_{0012} & - & 3\gamma_{0021} \\
       & - & 3\gamma_{0022} & - & 3\gamma_{0102} & - & 3\gamma_{0110} & - & 3\gamma_{0111} & - & 6\gamma_{0112} \\
       & - & 3\gamma_{0122} & - & 3\gamma_{0210} & - & 3\gamma_{0211} & - & 3\gamma_{0212} & - & 3\gamma_{0220} \\
       & - & 3\gamma_{0221} & - & 3\gamma_{0222} & - & 3\gamma_{1000} & - & 3\gamma_{1001} & - & 3\gamma_{1002} \\
       & - & 3\gamma_{1020} & - & 3\gamma_{1021} & - & 3\gamma_{1022} & - & 3\gamma_{1100} & - & 3\gamma_{1102} \\
       & - & 3\gamma_{1110} & - & 3\gamma_{1112} & - & 3\gamma_{1120} & - & 3\gamma_{1122} & - & 3\gamma_{1200} \\
       & - & 3\gamma_{1210} & - & 6\gamma_{1220} & - & 3\gamma_{1221} & - & 3\gamma_{1222} & - & 3\gamma_{2000} \\
       & - & 6\gamma_{2001} & - & 3\gamma_{2002} & - & 3\gamma_{2011} & - & 3\gamma_{2021} & - & 3\gamma_{2100} \\
       & - & 3\gamma_{2101} & - & 3\gamma_{2102} & - & 3\gamma_{2110} & - & 3\gamma_{2111} & - & 3\gamma_{2112} \\
       & - & 3\gamma_{2200} & - & 3\gamma_{2201} & - & 3\gamma_{2210} & - & 3\gamma_{2211} & - & 3\gamma_{2220} \\
       & - & 3\gamma_{2221} , & & & & & & & &
     \end{array} \nonumber
  \end{displaymath}
  which according to Eq.~(\ref{eq_5}) is less than or equal to $0$ and
  greater or equal to $-6$. Please note that $\gamma$'s are all positive here.

\section{The Exact Value of Equality ${\mathbb E}$ \label{app_3}}
  To find the exact value of the Bell expression ${\mathbb E}$, i.e.
  Eq.~(\ref{eq_16}), let's start with 
  \begin{displaymath}
     {\mathbb E} = + P_{11}^{00} + P_{11}^{01} + P_{11}^{10}
       + P_{11}^{11} + P_{12}^{22} + P_{21}^{12} - P_{22}^{12} .
  \end{displaymath}
  Using appendix~\ref{app_1} to write $P$'s in terms of $\gamma$'s we get
  \begin{displaymath}
     \begin{array}{llllllllllll}
     {\mathbb E} =
       & + & \gamma_{0000} & + & \gamma_{0001} & + & \gamma_{0002} & + & \gamma_{0100} & + & \gamma_{0101} \\
       & + & \gamma_{0102} & + & \gamma_{0200} & + & \gamma_{0201} & + & \gamma_{0202} &   & \vspace*{3pt} \\
       & + & \gamma_{0010} & + & \gamma_{0011} & + & \gamma_{0012} & + & \gamma_{0110} & + & \gamma_{0111} \\
       & + & \gamma_{0112} & + & \gamma_{0210} & + & \gamma_{0211} & + & \gamma_{0212} &   & \vspace*{3pt} \\
       & + & \gamma_{1000} & + & \gamma_{1001} & + & \gamma_{1002} & + & \gamma_{1100} & + & \gamma_{1101} \\
       & + & \gamma_{1102} & + & \gamma_{1200} & + & \gamma_{1201} & + & \gamma_{1202} &   & \vspace*{3pt} \\
       & + & \gamma_{1010} & + & \gamma_{1011} & + & \gamma_{1012} & + & \gamma_{1110} & + & \gamma_{1111} \\
       & + & \gamma_{1112} & + & \gamma_{1210} & + & \gamma_{1211} & + & \gamma_{1212} &   & \vspace*{3pt} \\
       & + & \gamma_{2002} & + & \gamma_{2012} & + & \gamma_{2022} & + & \gamma_{2102} & + & \gamma_{2112} \\
       & + & \gamma_{2122} & + & \gamma_{2202} & + & \gamma_{2212} & + & \gamma_{2222} &   & \vspace*{3pt} \\
       & + & \gamma_{0120} & + & \gamma_{0121} & + & \gamma_{0122} & + & \gamma_{1120} & + & \gamma_{1121} \\
       & + & \gamma_{1122} & + & \gamma_{2120} & + & \gamma_{2121} & + & \gamma_{2122} &   & \vspace*{3pt} \\
       & - & \gamma_{0202} & - & \gamma_{0212} & - & \gamma_{0222} & - & \gamma_{1202} & - & \gamma_{1212} \\
       & - & \gamma_{1222} & - & \gamma_{2202} & - & \gamma_{2212} & - & \gamma_{2222} . & & \vspace*{3pt} \\
     \end{array} \nonumber
  \end{displaymath}
  Simplifying the above equation would yield
  \begin{displaymath}
     \begin{array}{llllllllllll}
     {\mathbb E} =
       & + & \gamma_{0000} & + & \gamma_{0001} & + & \gamma_{0002} & + & \gamma_{0010} & + & \gamma_{0011} \\
       & + & \gamma_{0012} & + & \gamma_{0100} & + & \gamma_{0101} & + & \gamma_{0110} & + & \gamma_{0111} \\
       & + & \gamma_{0120} & + & \gamma_{0121} & + & \gamma_{0200} & + & \gamma_{0201} & + & \gamma_{0202} \\
       & + & \gamma_{0210} & + & \gamma_{0211} & + & \gamma_{0212} & + & \gamma_{1000} & + & \gamma_{1001} \\
       & + & \gamma_{1002} & + & \gamma_{1010} & + & \gamma_{1011} & + & \gamma_{1012} & + & \gamma_{1100} \\
       & + & \gamma_{1101} & + & \gamma_{1110} & + & \gamma_{1111} & + & \gamma_{1120} & + & \gamma_{1121} \\
       & + & \gamma_{1200} & + & \gamma_{1201} & + & \gamma_{1202} & + & \gamma_{1210} & + & \gamma_{1211} \\
       & + & \gamma_{1212} & + & \gamma_{2002} & + & \gamma_{2012} & + & \gamma_{2022} & + & \gamma_{2120} \\
       & + & \gamma_{2121} & + & \gamma_{2122} & + & \gamma_{2202} & + & \gamma_{2212} & + & \gamma_{2222} .
     \end{array} \nonumber
  \end{displaymath}
  With similar procedure, Eq.~(\ref{eq_17}), i.e.
  \begin{displaymath}
    {\mathbb E}_c = + P_{11}^{02} + P_{11}^{12} + P_{12}^{20}
       + P_{12}^{21} - P_{21}^{12} + P_{22}^{12} ,
  \end{displaymath}
  in terms of $\gamma$'s would become
  \begin{displaymath}
    \begin{array}{llllllllllll}
    {\mathbb E}_c =
     & + & \gamma_{0020} & + & \gamma_{0021} & + & \gamma_{0022} & + & \gamma_{0120} & + & \gamma_{0121} \\
     & + & \gamma_{0122} & + & \gamma_{0220} & + & \gamma_{0221} & + & \gamma_{0222} &   & \vspace*{3pt} \\
     & + & \gamma_{1020} & + & \gamma_{1021} & + & \gamma_{1022} & + & \gamma_{1120} & + & \gamma_{1121} \\
     & + & \gamma_{1122} & + & \gamma_{1220} & + & \gamma_{1221} & + & \gamma_{1222} &   & \vspace*{3pt} \\
     & + & \gamma_{2000} & + & \gamma_{2010} & + & \gamma_{2020} & + & \gamma_{2100} & + & \gamma_{2110} \\
     & + & \gamma_{2120} & + & \gamma_{2200} & + & \gamma_{2210} & + & \gamma_{2220} &   & \vspace*{3pt} \\
     & + & \gamma_{2001} & + & \gamma_{2011} & + & \gamma_{2021} & + & \gamma_{2101} & + & \gamma_{2011} \\
     & + & \gamma_{2121} & + & \gamma_{2201} & + & \gamma_{2211} & + & \gamma_{2221} &   & \vspace*{3pt} \\
     & - & \gamma_{0120} & - & \gamma_{0121} & - & \gamma_{0122} & - & \gamma_{1120} & - & \gamma_{1121} \\
     & - & \gamma_{1122} & - & \gamma_{2120} & - & \gamma_{2121} & - & \gamma_{2122} &   & \vspace*{3pt} \\
     & + & \gamma_{0202} & + & \gamma_{0212} & + & \gamma_{0222} & + & \gamma_{1202} & + & \gamma_{1212} \\
     & + & \gamma_{1222} & + & \gamma_{2202} & + & \gamma_{2212} & + & \gamma_{2222} . &   & \vspace*{3pt} \\
     \end{array} \nonumber
  \end{displaymath}
  And after simplifying we obtain
  \begin{displaymath}
     \begin{array}{llllllllllll}
     {\mathbb E}_c =
       & + & \gamma_{0020} & + & \gamma_{0021} & + & \gamma_{0022} & + & \gamma_{0102} & + & \gamma_{0112} \\
       & + & \gamma_{0122} & + & \gamma_{0220} & + & \gamma_{0221} & + & \gamma_{0222} & + & \gamma_{1020} \\
       & + & \gamma_{1021} & + & \gamma_{1022} & + & \gamma_{1102} & + & \gamma_{1112} & + & \gamma_{1122} \\
       & + & \gamma_{1220} & + & \gamma_{1221} & + & \gamma_{1222} & + & \gamma_{2000} & + & \gamma_{2001} \\
       & + & \gamma_{2010} & + & \gamma_{2011} & + & \gamma_{2020} & + & \gamma_{2021} & + & \gamma_{2100} \\
       & + & \gamma_{2101} & + & \gamma_{2102} & + & \gamma_{2110} & + & \gamma_{2111} & + & \gamma_{2112} \\
       & + & \gamma_{2200} & + & \gamma_{2201} & + & \gamma_{2210} & + & \gamma_{2211} & + & \gamma_{2220} \\
       & + & \gamma_{2221} . & & & & & & & &
     \end{array} \nonumber
  \end{displaymath}
  Adding ${\mathbb E}$ and ${\mathbb E}_c$ would result
  \begin{displaymath}
     \begin{array}{lllllllllllll}
     {\mathbb E} + {\mathbb E}_c
       & = & + & \gamma_{0000} & + & \gamma_{0001} & + & \gamma_{0002} & + & \gamma_{0010} & + & \gamma_{0011} \\
       &   & + & \gamma_{0012} & + & \gamma_{0020} & + & \gamma_{0021} & + & \gamma_{0022} & + & \gamma_{0100} \\
       &   & + & \gamma_{0101} & + & \gamma_{0102} & + & \gamma_{0110} & + & \gamma_{0111} & + & \gamma_{0112} \\
       &   & + & \gamma_{0120} & + & \gamma_{0121} & + & \gamma_{0122} & + & \gamma_{0200} & + & \gamma_{0201} \\
       &   & + & \gamma_{0202} & + & \gamma_{0210} & + & \gamma_{0211} & + & \gamma_{0212} & + & \gamma_{0220} \\
       &   & + & \gamma_{0221} & + & \gamma_{0222} & + & \gamma_{1000} & + & \gamma_{1001} & + & \gamma_{1002} \\
       &   & + & \gamma_{1010} & + & \gamma_{1011} & + & \gamma_{1012} & + & \gamma_{1020} & + & \gamma_{1021} \\
       &   & + & \gamma_{1022} & + & \gamma_{1100} & + & \gamma_{1101} & + & \gamma_{1102} & + & \gamma_{1110} \\
       &   & + & \gamma_{1111} & + & \gamma_{1112} & + & \gamma_{1120} & + & \gamma_{1121} & + & \gamma_{1122} \\
       &   & + & \gamma_{1200} & + & \gamma_{1201} & + & \gamma_{1202} & + & \gamma_{1210} & + & \gamma_{1211} \\
       &   & + & \gamma_{1212} & + & \gamma_{1220} & + & \gamma_{1221} & + & \gamma_{1222} & + & \gamma_{2000} \\
       &   & + & \gamma_{2001} & + & \gamma_{2002} & + & \gamma_{2010} & + & \gamma_{2011} & + & \gamma_{2012} \\
       &   & + & \gamma_{2020} & + & \gamma_{2021} & + & \gamma_{2022} & + & \gamma_{2100} & + & \gamma_{2101} \\
       &   & + & \gamma_{2102} & + & \gamma_{2110} & + & \gamma_{2111} & + & \gamma_{2112} & + & \gamma_{2120} \\
       &   & + & \gamma_{2121} & + & \gamma_{2122} & + & \gamma_{2200} & + & \gamma_{2201} & + & \gamma_{2202} \\
       &   & + & \gamma_{2210} & + & \gamma_{2211} & + & \gamma_{2212} & + & \gamma_{2220} & + & \gamma_{2221} \\
       &   & + & \gamma_{2222} & & & & & & & & \\
       & = &  1. & & & & & & & &
     \end{array} \nonumber
  \end{displaymath}
  The last equality holds due to Eq.~(\ref{eq_1}). Finally as a result of
  positivity of $\gamma$'s, we conclude that
  \be
    |{\mathbb E}| + |{\mathbb E}_c| = 1. \nonumber
  \ee
  So if according to quantum theory the value of $|{\mathbb E}_c|$ is known,
  then the value of $|{\mathbb E}|$ would also be known. And in the special case
  that ${\mathbb E}$ is positive then the exact value of ${\mathbb E}$ would be
  specified.

\end{document}